# $QED_3$ theory of underdoped high temperature superconductors II: the quantum critical point

Dominic J. Lee and Igor F. Herbut

*Department of Physics, Simon Fraser University, Burnaby, British Columbia, Canada V5A 1S6*

We study the effect of gapless quasiparticles in a d-wave superconductor on the $T = 0$ end point of the Kosterlitz-Thouless transition line in underdoped high-temperature superconductors. Starting from a lattice model that has gapless fermions coupled to 3D XY phase fluctuations of the superconducting order parameter, we propose a continuum field theory to describe the quantum phase transition between the d-wave superconductor and the spin-density-wave insulator. Without fermions the theory reduces to the standard Higgs scalar electrodynamics (HSE), which is known to have the critical point in the inverted XY universality class. Extending the renormalization group calculation for the HSE to include the coupling to fermions, we find that the qualitative effect of fermions is to increase the portion of the space of coupling constants where the transition is discontinuous. The critical exponents at the stable fixed point vary continuously with the number of fermion fields $N$, and we estimate the correlation length exponent ($\nu = 0.65$) and the vortex field anomalous dimension ($\eta_\Phi = -0.48$) at the quantum critical point for the physical case $N = 2$. The stable critical point in the theory disappears for the number of Dirac fermions $N > N_c$, with $N_c \approx 3.4$ in our approximation. We discuss the relationship between the superconducting and the chiral (SDW) transitions, and point to some interesting parallels between our theory and the Thirring model.

## I. INTRODUCTION

All high-temperature superconductors are well established to have d-wave symmetry of their order parameter [1], [2], and therefore to posses gapless fermionic excitations near the four points in the momentum space where the order parameter vanishes. Another distinguishing feature of high-temperature superconductors is that, to a very good approximation, they may be considered to be quasi-two-dimensional when underdoped [3]. This leads one to expect the superconducting transition at finite temperatures ($T \neq 0$) in the underdoped regime to be of the Kosterlitz-Thouless type [4]. Indeed, there is mounting experimental support in favour of free vortices near and above the superconducting critical temperature [5], [6]. Building on these two observations, one of us [7] has recently shown that in the minimal theory that couples vortex and quasiparticle degrees of freedom [8] the result of phase disordering in a d-wave superconductor (dSC) at $T = 0$ is nothing but the antiferromagnetic (more precisely, the incommensurate spin-density wave (SDW)) insulator, the phase that is believed to be neighbouring the dSC in the phase diagram of all high-temperature superconductors. This result follows from realizing that integration of the unbound vortex degrees of freedom yields the theory for the low-energy fermions that is closely related to the two-component three dimensional quantum electrodynamics ($QED_3$) [8], [7] in which the dynamical breaking [9] of an approximate "chiral" symmetry translates into SDW ordering [10].

One can show [11] that gapless quasiparticles do not change the Kosterlitz-Thouless universality class of the $T \neq 0$ superconducting transition, as one could also expect on quite general grounds. At $T = 0$, on the other hand, the effect of gapless fermionic excitations on the order parameter is often less benign, and may change the universality class, or even the order, of the phase transition [12]. In this paper we study the quantum ($T = 0$) phase transition in the theory of coupled gapless quasiparticles and vortex-loop excitations in the phase of the superconducting order parameter. On the insulating side, i. e. when the vortex-loops are unbound, our theory reduces to the effective $QED_3$ description of underdoped cuprates [7], [8]. Assuming that the dynamics of the superconducting phase at $T = 0$ is governed by the three dimensional (3D) XY model, we show that the effect of gapless quasiparticles is then to introduce an *additional* massless gauge field into the Higgs scalar electrodynamics (HSE), which represents the *dual* theory for vortex-loop excitations. The effect of this extra gauge-field on the quantum critical point is controlled by the number of fermionic fields $N$, and for $N = 0$ we recover the results for standard HSE [13]. Using the approximate fixed dimension renormalization group (RG) equations for the HSE, known to be qualitatively and even quantitatively reliable when $N = 0$ [13], we show that there exists a critical number of fermionic fields $N_c \approx 3.4$, above which there are no longer stable fixed points in the theory, and the flow in the space of coupling constants is always towards the region with a negative quartic term coupling. We interpret these runaway flows as the fluctuation induced first-order transition. Similar conclusion is reached in a more controlled calculation in $4-\epsilon$ dimensions, where we find that the critical number of complex components of the scalar field above which a second order phase transition can occur [14] *increases* when fermions are present. This suggests that gapless quasiparticles tend to destabilize the inverted XY phase transition in favour of a first order one. For the physical case $N = 2$, however, we still find a critical point within our calculation, and we assume it controls the quantum phase transition for a strongly type-II HSE. Its critical exponents are estimated to be $\nu = 0.65$ and $\eta_\Phi = -0.48$.

The paper is organised as follows. In the next section we define the lattice theory one can use to model the system of gapless fermionic excitations coupled to vortex loops. In the sec. III we propose the continuum description of the lattice theory and discuss some of its features and possible problems. We then proceed to extend the fixed dimension RG calculation of ref. [13] onto the present problem in sec. IV. A supporting computation in $4 - \epsilon$ dimensions is provided in sec. V. In sec. VI we discuss a simpler version of our theory. Conclusion and discussion of our results are presented in sec. VII. Technical details are given in Appendices.

## II. THE LATTICE MODEL

We begin with the simplest lattice model that has gapless fermions coupled to the XY variables:

$$Z = \int \prod_x (d\bar{\chi}_i(x) d\chi_i(x) d\mathbf{a}(x)) \quad (1)$$

$$\sum_{\mathbf{m}_A(x),\mathbf{m}_B(x)}' \exp-(S_1[\chi,\mathbf{a}] + S_2[\mathbf{a},\mathbf{m}_A,\mathbf{m}_B]),$$

where

$$S_1[\chi,\mathbf{a}] = \frac{1}{2} \sum_{i=1}^{N/2} \sum_{x,x',\mu} \quad (2)$$

$$\eta_\mu(x) \bar{\chi}_i(x) [\delta_{x',x+\hat{\mu}} e^{ia_\mu(x)} - \delta_{x',x-\hat{\mu}} e^{ia_\mu(x-\hat{\mu})}] \chi_i(x'),$$

and

$$S_2[\mathbf{a},\mathbf{m}_A,\mathbf{m}_B] = \sum_x \{\frac{1}{8K} [\nabla \times (\mathbf{m}_A(x) + \mathbf{m}_B(x))]^2 \quad (3)$$

$$+ i\mathbf{a}(x) \cdot [\nabla \times (\mathbf{m}_A(x) - \mathbf{m}_B(x))]\}.$$

Here $x$ labels the sites of a three dimensional quadratic lattice, $\mu = 0,1,2$, and $\nabla \times$ denotes the lattice curl. Prime on the sum in (1) means that $\nabla \cdot \mathbf{m}_{A,B} = 0$, where $\nabla$ is the lattice gradient, and $\mathbf{m}_{A,B}$ are integer vector variables. $\eta_\mu(x)$ are the standard Kawamoto-Smit phases ($\eta_1(x) = 1$, $\eta_2(x) = (-1)^{x_1}$, $\eta_0(x) = (-1)^{x_1+x_2}$) introduced to ensure relativistic covariance in the continuum limit [15], [16]. $\chi_i(x)$ are one component Grasmann variables. To be specific, we are using staggered fermions, and will, therefore, consider $N$ to be even.

Detailed derivation of the above partition function may be found in [7]. In the continuum limit it describes $N$ four-component Dirac fermions [15], which represent the neutral gapless spin-$1/2$ excitations (spinons) that can be identified in the d-wave superconducting state, coupled to XY degrees of freedom. Two species of integers ($A$ and $B$) are the result of making the Franz-Tešanović transformation on the original electrons [8], in which spin up and spin down electrons need to be transformed differently.

Notice that in the absence of fermions (when $N = 0$) the integration over the gauge-field $\mathbf{a}$ enforces the constraint $\mathbf{m}_A(x) = \mathbf{m}_B(x)$, so that the partition function reduces to the well-known current-loop representation of the XY model [17]. The parameter $K$ is the XY phase stiffness, which may be understood as proportional to doping [18], at least in the low doping regime of high-temperature superconductors.

The problem we wish to address here is how the XY universality class is modified by the presence of gapless fermionic excitations in the d-wave superconductor. The case which is of particular interest is when $N = 2$, corresponding to four nodes of the d-wave order parameter in a single-layer superconductor.

The lattice theory (1) can in principle be simulated on a computer, and the nature of the phase transition as the parameter $K$ is varied studied that way. In this paper we will make some further transformations that will enable us to suggest a continuum field theory which should have the phase transition in the same universality class as the lattice model. For that purpose we first rewrite the above partition function by using the standard dual (vortex loop) variables $\mathbf{l}_A$ and $\mathbf{l}_B$ as

$$Z = \lim_{y \to 0} \int \prod_x (d\bar{\chi}_i(x) d\chi_i(x) d\mathbf{a}(x) d\mathbf{A}_A(x) d\mathbf{A}_B(x)) \quad (4)$$

$$\sum_{\mathbf{l}_A(x),\mathbf{l}_B(x)}' \exp-\{S_1[\chi,\mathbf{a}] + S_2[\mathbf{a},\mathbf{A}_A,\mathbf{A}_B] +$$

$$\sum_x i2\pi (\mathbf{l}_A(x) \cdot \mathbf{A}_A(x) + \mathbf{l}_B(x) \cdot \mathbf{A}_B(x))$$

$$+ y(\mathbf{l}_A^2(x) + \mathbf{l}_B^2(x))\}.$$

When the "chemical potential" for the vortex loops $y = 0$ the summation over integers $\mathbf{l}_A(x)$ and $\mathbf{l}_B(x)$ forces $\mathbf{A}_A(x)$ and $\mathbf{A}_B(x)$ to take strictly integer values, and one recovers the previous expression for the partition function. The approximation we will make is to take $y$ small but finite. This approximation is often made in studies of the 2D [19] and the 3D [20] XY model, and is known not to change the universality class of the transition in that case [21]. In present case the innocence of such an approximation is less clear. First, when $y = 0$ the action is symmetric under the transformation $a_\mu(x) \to a_\mu(x) + 2\pi n_\mu(x)$ with $n_\mu(x)$ integer, and the summation over integers $\mathbf{m}_{A,B}$ would in principle yield a *compact* gauge theory. This symmetry is broken when $y \neq 0$, and one therefore replaces in this way a compact gauge theory with a non-compact one. The difference between the two, when the gauge-field is coupled to matter, is at present far from clear [22], and we will unfortunately have nothing further to add on this issue. Second, in the quenched limit $N = 0$ the partition function for $y \neq 0$ does not exactly reduce to the XY model any longer. Consequently, our continuum dual theory, which will assume $y \neq 0$ at $N = 0$, will not automatically be in the inverted XY universality class, as it should in the absence of fermions. We will, however, propose a way how to deal with this problem in the next section.

With these two caveats in mind we can proceed towards the continuum limit by observing that the last expression is up to the Villain transformation [23] equivalent to

$$Z = \lim_{y \to 0} \int \prod_x (d\bar{\chi}_i(x) d\chi_i(x) d\mathbf{a}(x) d\mathbf{A}_A(x) d\mathbf{A}_B(x) \quad (5)$$
$$d\theta_A(x) d\theta_B(x)) \exp{-(S_1[\chi, \mathbf{a}] + S_2[\mathbf{a}, \mathbf{A}_A, \mathbf{A}_B]}$$
$$-\frac{1}{2y} \sum_x \{\cos(\theta_A(x) - \theta_A(x+\hat{\nu}) - 2\pi A_{A,\nu}(x))$$
$$+ \cos(\theta_B(x) - \theta_B(x+\hat{\nu}) - 2\pi A_{B,\nu}(x))\}.$$

Relaxing the $y \to 0$ limit enables one to recognise the continuum limit of the lattice theory given in Eq. (5), which is what we do in the next section.

## III. DUAL CONTINUUM THEORY

We write down a generalized continuum theory using n-component complex scalar fields. This model will be in the same universality class as the lattice ($y \neq 0$) model in Eq. (5) for $n = 1$, and $b_1 = b_2 \gg e^2$ (strong type-II limit):

$$\mathcal{L} = \mathcal{L}_h + \mathcal{L}_f + \mathcal{L}_c,$$
$$\mathcal{L}_f = i \sum_{j=1}^N \bar{\Psi}_j \left(\slashed{\partial} - i\lambda \slashed{a}\right) \Psi_j,$$
$$\mathcal{L}_h = |\left(\nabla - i\mathbf{A}_+ - i\mathbf{A}_-\right) \mathbf{\Phi}_1|^2 + t_0 |\mathbf{\Phi}_1|^2$$
$$+ |\left(\nabla - i\mathbf{A}_+ + i\mathbf{A}_-\right) \mathbf{\Phi}_2|^2 + t_0 |\mathbf{\Phi}_2|^2$$
$$+ \frac{b_1}{4} \left[ \left(|\mathbf{\Phi}_1|^2\right)^2 + \left(|\mathbf{\Phi}_2|^2\right)^2 \right] + \frac{b_2}{2} |\mathbf{\Phi}_1|^2 |\mathbf{\Phi}_2|^2$$
$$+ \frac{(\nabla \times \mathbf{A}_+)^2}{2e^2},$$
$$\mathcal{L}_c = \frac{i\mathbf{a} \cdot \nabla \times \mathbf{A}_-}{\pi}. \quad (6)$$

$\Psi_j$ are the *four-component* spinon fields [7] that represent the gapless quasiparticle excitations, and which are the continuum limit of the $\chi_i$ lattice fermions. $\mathbf{A}_-$ and $\mathbf{A}_+$ are the continuum versions of $\pi(\mathbf{A}_A - \mathbf{A}_B)$ and $\pi(\mathbf{A}_A + \mathbf{A}_B)$. The two scalar fields $\mathbf{\Phi}_1$ and $\mathbf{\Phi}_2$ are the "soft-spin" representation of the dual angles $\theta_A$ and $\theta_B$ in the continuum limit. There are two couplings that describe the interactions between $\mathbf{\Phi}_1$ and $\mathbf{\Phi}_2$ fields, $b_1$ and $b_2$. We have also introduced the parameter $\lambda$ to control the strength of the coupling to spinons. The physical case corresponds to $\lambda = 1$. Finally, tuning the parameter $K$ in the lattice theory should correspond to varying $t_0$ in the Eq. (6).

At the *mean-field* level the above continuum theory has a *single* phase transition from the insulating phase of unbound vortex loops $\langle \mathbf{\Phi_1} \rangle = \langle \mathbf{\Phi_2} \rangle = \langle \mathbf{\Phi} \rangle \neq 0$, when $t_0 < 0$, to the superconducting phase (bound vortex loops) $\langle \mathbf{\Phi_1} \rangle = \langle \mathbf{\Phi_2} \rangle = 0$, when $t_0 > 0$. To understand the fate of fermionic degrees of freedom in both phases one needs to integrate out the gauge fields $\mathbf{A}_-$ and $\mathbf{A}_+$:

1) In the superconducting phase $\mathbf{A}_+$ is decoupled from $\mathbf{a}$, and therefore decoupled from spinons. Integration over $\mathbf{A}_-$ gives a constraint $\nabla \times \mathbf{a} = 0$, and therefore the spinons are free. This corresponds to sharp quasiparticle excitations in the dSC.

2) In the insulating phase there is a Meissner effect for both $\mathbf{A}_+$ and $\mathbf{A}_-$, and the term $\sim 2|\langle\mathbf{\Phi}\rangle|^2(\mathbf{A}_+^2 + \mathbf{A}_-^2)$ appears in the action. $\mathbf{A}_+$ is still decoupled from $\mathbf{a}$, but the integration over $\mathbf{A}_-$ now yields

$$\mathcal{L} \to \sum_{j}^{N} \bar{\Psi}_j \left(\slashed{\partial} - i\lambda \slashed{a}\right) \Psi_j + \frac{(\nabla \times \mathbf{a})^2}{32\pi^2 |\langle\mathbf{\Phi}\rangle|^2}. \quad (7)$$

This is three dimensional quantum electrodynamics ($QED_3$) [7], [8]. For $N < N_c \approx 3$ the "chiral" symmetry [9] is spontaneously broken, and the "chiral condensate" $\langle\bar{\psi}\psi\rangle \propto |\langle\mathbf{\Phi}\rangle|^2$ appears. Going back to the electronic origin of the $QED_3$ one finds that the chiral condensate is precisely the SDW order parameter [7].

The continuum model has three local U(1) gauge symmetries, one for each of the three gauge fields. These symmetries should insure that the theory is renormalizable, and that $\mathbf{A}_-$ and $\mathbf{A}_+$, for example, do not acquire a mass in the superconducting ($\langle\mathbf{\Phi}\rangle = 0$) phase. To preserve this gauge symmetry, however, one needs to keep two types of the fluctuating $\mathbf{\Phi}$ fields: if one would simply set $\mathbf{\Phi}_1 = \mathbf{\Phi}_2$ in Eq. (6) the model would lose the $U(1)$ symmetry associated with the $\mathbf{A}_-$ field. This would lead to a theory that is not renormalizable; the $\mathbf{A}_-$ would becomes massive in the superconducting phase, as the Ward identities which usually prevent this from happening would be no longer respected.

In absence of fermions (i. e. at $N = 0$, or $\lambda = 0$) and when $b_1 = b_2$, if $\mathbf{a}$ is integrated out, $\mathbf{A}_-$ becomes constrained to be zero. The model simply reduces then to 3D HSE with *two* n-component complex fields. As discussed in the previous section, this doubling of degrees of freedom should be considered to be an artifact of our relaxation of the $y \to 0$ limit in the underlying lattice theory. We therefore impose that for $N = 0$ our calculation reduces to the HSE with a *single* complex scalar field by formally taking the number of complex components for each field in our model to be $n = 1/2$, not $n = 1$. As we will be performing perturbative calculations, $n$ will determine only the coefficients in our expansions, and therefore for $n = 1/2$ and $N = 0$ those will be then exactly the same as in the HSE. This computational device will allow us to asses the effect of the coupling to gapless fermions.

If we were to set $b_1 = b_2$ in the Eq. (6) we would have a global $U(2n)$ symmetry associated with the $2n$ $\mathbf{\Phi}$ fields, when $\mathbf{A}_- = 0$. However, the coupling to the gauge field $\mathbf{A}_-$ breaks such a $U(2n)$ invariance. Hence, if initially $b_1 = b_2$, quantum corrections would cause $b_1$ and $b_2$ to be no longer equal in the effective action. To consistently renormalize the theory we therefore need to keep track of both couplings in the intermediate stages of the calculation. Only at the very end will we impose the constraint that $\langle \mathbf{\Phi}_1 \rangle = \langle \mathbf{\Phi}_2 \rangle$ in the effective action, once the quantum corrections have been computed.

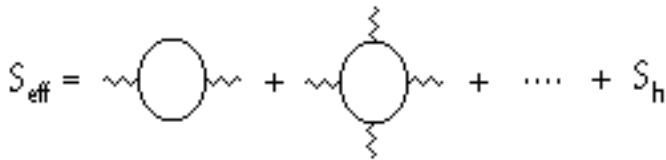

FIG. 1: Diagram showing non-vanishing terms that contribute to the effective theory obtained by integrating out the fermion fields. The smooth solid lines represent fermion propagators. The wavy lines represent gauge fields.

## IV. RENORMALIZATION GROUP IN 3D

We proceed to study the continuum theory in Eq. (6) by using one-loop renormalization group at fixed dimension. Although such a calculation is essentially uncontrolled, it has been previously used successfully in [13] by appropriately choosing the renormalization conditions at the critical point $t = 0$. We will find that by increasing $\lambda$ one progressively destabilizes the second order phase transition in favour of the first order one. Above certain $\lambda = \lambda_c$ the phase transition becomes first order for all values of the couplings $e$, $b_1$ and $b_2$. To show that this may not simply be an artifact of our approximation, in the next section we reach a similar conclusion by using the minimal subtraction scheme in $4 - \epsilon$ dimensions.

We begin by integrating out the fermionic fields and expand out in powers of the vortex gauge field $\mathbf{a}$. Such an expansion is shown in Fig. 1. For $\lambda \ll 1$ we may retain only the first non-vanishing term in such an expansion, which is proportional to the polarization tensor, $\Pi_{\mu\nu}$. The $\mathbf{a}$ field may then be integrated out exactly, and we are left with an effective theory

$$S_{eff} = \int d^3x \left[ \frac{(\nabla \times \mathbf{A}_-(x))^2}{2|\nabla|\bar{\lambda}^2} + \mathcal{L}_h \right], \qquad (8)$$

where $\bar{\lambda}^2 = \frac{\lambda^2 N \pi^2}{8}$ is an effective "charge" associated with the $\mathbf{A}_-$ field, introduced by fermions. In a less approximate treatment one would also find the terms of higher order in the gauge-field $\mathbf{A}_-$, as well as a renormalization of the coupling in the $\mathbf{A}_-^2$ term. The effect of including these higher order corrections, will be to renormalize the effective charge $\bar{\lambda}$, so that $\bar{\lambda}$ will be a power series in both $\lambda$ and $1/N$, with the value of $\bar{\lambda}$ given above as the leading order term. For the sake of simplicity, we will however neglect these higher order corrrections.

Next, we compute the polarization tensors $2n\bar{\lambda}^2 \chi_{\mu\nu}^-$ and $2ne^2 \chi_{\mu\nu}^+$, for the $\mathbf{A}_-$, $\mathbf{A}_+$ fields, respectively, to one loop order. It is easy to show that $\chi_{\mu\nu}^-=\chi_{\mu\nu}^+$ to all orders in perturbation theory. We give an explicit expression for $\chi_{\mu\nu}^+$ in Appendix A. Another important observation is that there is no mixing between $\mathbf{A}_-$ and $\mathbf{A}_+$, again to all orders of perturbation theory. This is because $\mathbf{A}_-$ couples to $\mathbf{\Phi}_2$ with opposite charge to the coupling of $\mathbf{A}_-$ to $\mathbf{\Phi}_1$, so contributions from $\mathbf{\Phi}_1$ fields to the mixing term always cancel those of $\mathbf{\Phi}_2$ fields. Finally, the $\mathbf{\Phi}_1$ propagator is exactly equal to the $\mathbf{\Phi}_2$ propagator. In Appendix A we provide the expressions for the inverse $\mathbf{\Phi}$ propagator $\Gamma^{(2)}$. Contributions to $\chi_{\mu\nu}$ and $\Gamma^{(2)}$ are shown diagrammatically in Fig. 2.

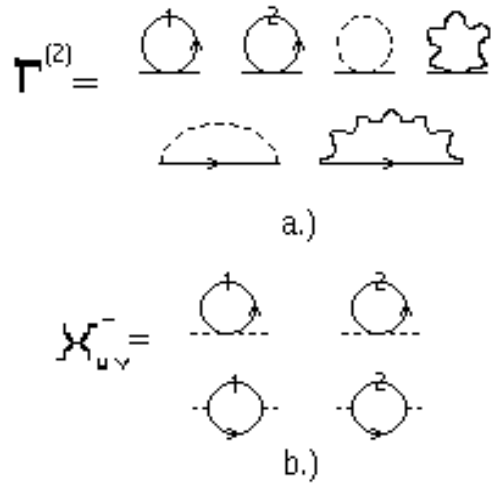

FIG. 2: Diagrammatic contributions to $\chi_{\mu\nu}^-$ and $\Gamma^{(2)}$. The smooth solid lines with numbers in them represent $\mathbf{\Phi}$-field propagators, the number 1 corresponding to $\mathbf{\Phi}_1$-field propagators, etc. The broken lines represent the propagators for $\mathbf{A}_-$, and the wavy lines for $\mathbf{A}_+$.

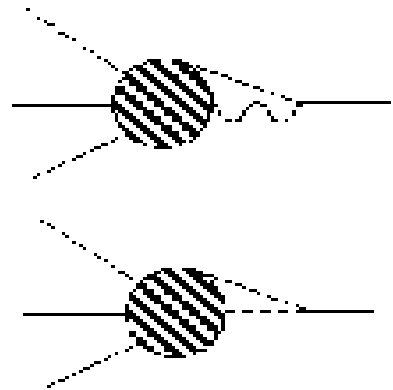

FIG. 3: Terms that do not contribute to the RG flow for $\Gamma^{(4)}_{1,1}$ and $\Gamma^{(4)}_{1,2}$. The cross hatched circle represents any allowed combination $\mathbf{\Phi}$, $\mathbf{A}_-$, and $\mathbf{A}_+$ propagators.

We then consider the four point vertex functions for the $\mathbf{\Phi}$ fields. For renormalization group purposes, we do not need to include all the terms that appear to one loop order. In the customary Landau (transverse) gauge, we adopt here, terms with the graphs of the form given in Fig. 3, do not affect the critical behaviour [24], since they vanish for finite $t$ when $p \to 0$. We shall denote the four point vertices for $\mathbf{\Phi}_1-\mathbf{\Phi}_1$ scattering and $\mathbf{\Phi}_2-\mathbf{\Phi}_2$ by $\Gamma^{(4)}_{1,1}$ and $\Gamma^{(4)}_{2,2}$, respectively. We shall use $\Gamma^{(4)}_{1,2}$, to denote the four point vertex of scattering between two $\mathbf{\Phi}_1$ and two $\mathbf{\Phi}_2$ fields.

We calculate these vertices at the symmetric point

$$\mathbf{p}_i.\mathbf{p}_j = p^2(4\delta_{ij} - 1),$$

where $\mathbf{p}_i$ are the momenta of the external legs. Some care must be taken when working in any fixed dimension other than 4. The reason for this is that, when considering $\Gamma^{(4)}_{1,1}$ at the symmetric point, two of the scattering channels (the t- and u-channels [25]) that contribute to the vertex have a total external momenta $p_1-p_2$, instead of $p_1+p_2$ for the other (s-) channel. This implies, at the symmetric point, $(p_1+p_2)^2$ gets replaced by $p^2$, as usual; but $(p_1-p_2)^2$ should be replaced by $2p^2$. So on calculating the loop integrals, an extra factor of $2^{(2-d/2)}$ appears in the terms that come from the t- and u-channels. The same consideration must be taken also when calculating the four point functions $\Gamma^{(4)}_{1,2}$ and $\Gamma^{(4)}_{2,2}$. Expressions for $\Gamma^{(4)}_{1,1}$, $\Gamma^{(4)}_{2,2}$ and $\Gamma^{(4)}_{1,2}$ at the symmetric point are given in Appendix A. For the purposes of illustration we show these terms in Fig. 4.

To derive expressions for the renormalized couplings, we now impose the following renormalization conditions in 3D

$$\begin{aligned}
\Gamma^{(2),R}(p) &= 0, \quad \frac{\partial \Gamma^{(2),R}(p)}{\partial p^2} = 1, \\
\Gamma^{(4),R}_{1,1}(p_1,p_2,p_3,p_4))_{SP} &= p\, b_{1,R}, \\
\Gamma^{(4),R}_{1,2}(p_1,p_2,p_3,p_4))_{SP} &= \frac{p b_{2,R}}{2}, \\
D^{+,R}(p) &= \frac{1}{q^2}\left(\delta_{\mu,\nu} - \frac{q_\mu q_\nu}{q^2}\right)_{q=p/c}, \\
D^{-,R}(p) &= \frac{1}{q}\left(\delta_{\mu,\nu} - \frac{q_\mu q_\nu}{q^2}\right)_{q=p/c}, \\
\Gamma^{(3)+,R}(p,0,p) &= p^{1/2} e_R, \\
\Gamma^{(3)-,R}(p,0,p) &= \bar{\lambda}_R.
\end{aligned} \quad (9)$$

Here, $\Gamma^{(3)+,R}$ and $\Gamma^{(3)-,R}$ are the renormalized $\mathbf{A}_+ - \mathbf{\Phi}_1 - \mathbf{\Phi}_1$ and $\mathbf{A}_- - \mathbf{\Phi}_1 - \mathbf{\Phi}_1$ vertices, respectively. As bare quantities, $\Gamma^{(3)-}$ and $\Gamma^{(3)+}$ take the values $\bar{\lambda}$ and $e$. Because of the Ward identities (given in Appendix A), we do not require an expression for $\Gamma^{(3)+}$ nor for $\Gamma^{(3)-}$, to one loop order, to determine renormalized charges; all that is required are the expressions for $D^+_{\mu,\nu}$ and $D^-_{\mu,\nu}$. Here, $c$ is an arbitrary scaling factor which we shall tune [13] so that the value of the Ginzburg-Landau parameter $b_1/4e^2$ at the tricritical point in standard HSE (i. e. when $\lambda = 0$) matches the known value. Using the renormalization conditions and the expressions in Appendix A we are able to write down the renormalized couplings for $n=1/2$, $d=3$

$$pb_{1,R} = b_1 - \left[\frac{7b_1^2}{32\sqrt{2}} + \frac{b_1^2}{16} + \frac{b_2^2}{32\sqrt{2}}\right]\frac{1}{p}$$

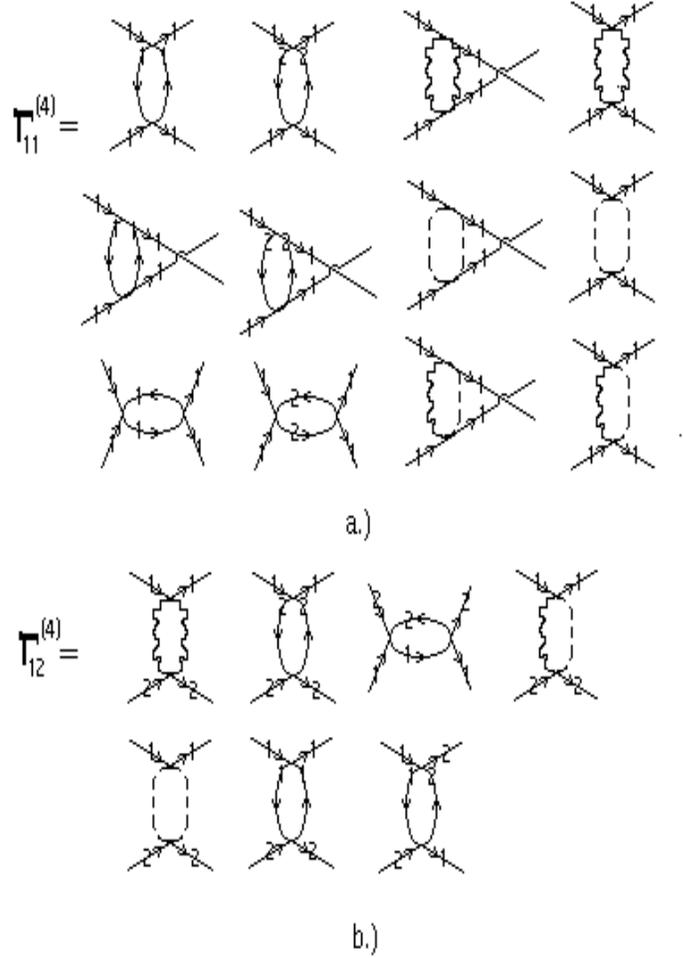

FIG. 4: Terms that contribute to renormalized $\Gamma^{(4)}_{1,1}$ and $\Gamma^{(4)}_{1,2}$.

$$\begin{aligned}
&+\frac{1}{2}\frac{e^2 b_1}{p} - \frac{3}{4\sqrt{2}}\frac{e^4}{p} + \left(\frac{2\bar{\lambda}^2 b_1}{\pi^2} - \frac{4e^2\bar{\lambda}^2}{\sqrt{2}\pi^2}\right)\frac{p^{-\epsilon'}}{\epsilon'} + \frac{4\bar{\lambda}^4 p}{\sqrt{2}\pi^2}, \\
&pb_{2,R} = b_2 - \left[\frac{b_2^2}{16}\left(1+\frac{1}{\sqrt{2}}\right) + \frac{3}{16\sqrt{2}}b_1 b_2\right]\frac{1}{p} \\
&+\frac{e^2 b_2}{2p} - \frac{3}{4\sqrt{2}}\frac{e^4}{p} + \left(\frac{2\bar{\lambda}^2 b_2}{\pi^2} + \frac{4e^2\bar{\lambda}^2}{\sqrt{2}\pi^2}\right)\frac{p^{-\epsilon'}}{\epsilon'} + \frac{4\bar{\lambda}^4 p}{\sqrt{2}\pi^2}, \\
&pe_R^2 = e^2 - \frac{e^4 c}{16p}, \\
&\bar{\lambda}_R^2 = \bar{\lambda}^2 - \frac{\bar{\lambda}^4 c}{16}
\end{aligned} \quad (10)$$

From the last line one can see that the coupling $\bar{\lambda}$ does not flow. In fact, the only reason why we found the $\bar{\lambda}^4$ term in $\bar{\lambda}_R$ is that we are working right at the critical point $t=0$; if this were not so the polarization would be proportional to $p^2$ and the coupling $\bar{\lambda}$ would not renormalize at all. This is because the Maxwell-like term for $\mathbf{A}_-$ in the Eq. (8) is non-analytic in small momentum, and therefore can not get renormalized [26]. We will therefore neglect the $\bar{\lambda}^4$ term in $\bar{\lambda}_R$, and consider $\bar{\lambda}$ to be an *exactly marginal* coupling hereafter.

In the Eq. (10) we introduced $\epsilon'$ as an infinitesimal quantity used to regularize the logarithmically singular terms that appear in the renormalized couplings. $\epsilon'$ is defined to be an infinitesimal shift in the scaling dimension of the $\mathbf{A}_-$ gauge field propagator, so in 3D

$$D^-_{\mu,\nu} = p^{-1-\epsilon'}\left(\delta_{\mu,\nu} - \frac{p_\mu p_\nu}{p^2}\right)$$

We choose to regularize this way, as opposed to dimensional regularization, due to fact that, when we generalise our model to arbitrary dimension (in the next section), we find that these logarithmic poles are present in all dimensions. The reader can easily convince him/herself that by setting $\bar{\lambda} = 0$ and $b_1 = b_2 = b$ one indeed finds the correct expressions for renormalized couplings for HSE to one loop order in $3D$. On rescaling, these expressions agree with [13], except for one difference; the $e^4$ term that contributes to $b_1$ and $b_2$, contains an extra piece, missed in [13]. In 4D this extra contribution can be neglected, as it is not singular; but in 3D it gives a $1/p$ contribution, so must be included in the $e^4$ term.

Before proceeding to the flow equations one needs to address the following technical issue: how does one treat the $\bar{\lambda}^4$ term in $b_{1,R}$ and $b_{2,R}$? If we were to blindly differentiate the expressions in Eq.(10), we would find the $\bar{\lambda}^4$ terms in both the beta functions for the $b_1$ and $b_2$ fields. But, power counting in 3D tells us that such terms are regular (i. e. not singular in the infrared limit) and should be discarded (together with all other regular contributions) from the beta functions. On the other hand, as we shall see, when close to four dimensions these terms should be included as they are as relevant as all the other terms. We think that the correct way to handle such terms when working directly in 3D, is to absorb these finite terms into bare values of the quartic couplings. That way they determine only the initial values of the flowing couplings (actually reduce $b_1$ and $b_2$) instead of entering the beta functions. At the end of this section we will briefly discuss what the effect of including such terms in the flow equations would be, to reassure the reader that our general conclusions would remain unchanged.

By making such a finite shift of $b_1$ and $b_2$, and then differentiating the remaining expressions we arrive at the flow equations,

$$p\frac{db_1}{dp} = -b_1 - \frac{1}{2}e^2 b_1 + \frac{3}{4\sqrt{2}}e^4 +$$
$$\left[b_1^2\left(\frac{7}{32\sqrt{2}} + \frac{1}{16}\right) + b_2^2 \frac{1}{32\sqrt{2}}\right] - \left(\frac{2\bar{\lambda}^2 b_1}{\pi^2} - \frac{4e^2\bar{\lambda}^2}{\sqrt{2}\pi^2}\right),$$
$$p\frac{db_2}{dp} = -b_2 - \frac{1}{2}e^2 b_2 + \frac{3}{4\sqrt{2}}e^4 +$$
$$\left[\frac{b_2^2}{16}\left(1 + \frac{1}{\sqrt{2}}\right) + \frac{3}{16\sqrt{2}}b_1 b_2\right] - \left(\frac{2\bar{\lambda}^2 b_2}{\pi^2} + \frac{4e^2\bar{\lambda}^2}{\sqrt{2}\pi^2}\right),$$
$$p\frac{de^2}{dp} = -e^2 + \frac{ce^4}{16},$$
(11)

| $\bar{\lambda}$ | $(b_-,\delta_-)$ | $(b_-,\delta_+)$ | $(b_+,\delta_-)$ | $(b_+,\delta_+)$ |
|---|---|---|---|---|
| 0 | (0.324,0) | (3.03,6.44) | (6.39,-5.53) | (5.78,0) |
| 0.4 | (0.347,-0.060) | (3.17,6.48) | (6.51,-5.78) | (5.83,0.289) |
| 1.0 | (0.455,-0.339) | (3.90,6.67) | (7.20,-6.92) | (6.12,1.63) |
| 1.5 | (0.582,-0.660) | (5.00,6.86) | (8.24,-8.41) | (6.55,3.40) |
| 2 | (0.711,-0.985) | (6.64,6.81) | (9.74,-10.3) | (7.04,5.91) |
| 2.5 | (0.824,-1.27) | N.A | (11.7,-12.7) | N.A |

TABLE I: Fixed points for various values of $\bar{\lambda}$.

where we omitted the subscript R.

First, we constrain the parameter $c$ in the same manner as in [13]: we set $\bar{\lambda} = 0$ and $b_1 = b_2$, calculate the position of the tricritical line in the flow diagram for the standard HSE, and then use the lattice value of the tricritical Ginzburg-Landau parameter $\kappa_c = b_1/(4e^2) = 0.42/\sqrt{2}$ [27] to fix $c$. This yields a value of c=17.4.

To analyse the flow equations it is convenient to define the new couplings $b = b_1$, $\delta = b_2 - b_1$. In Fig. 5 we show the numerically calculated flow for $e^2 = e^{*2} \equiv 16/17.4$, the fixed point value for $e$. We see in Fig. 5, for $\bar{\lambda} = 0$ and $\bar{\lambda} = 2$, that there are four fixed points. At $\lambda = 0$ we have the usual HSE fixed points at $\delta = 0$ [13]. These are the tricritical point $(b_-,\delta_-)$, which lies on the $\kappa_c$ line and has two repulsive eigenvalues in the $e^2 = e^{*2}$ plane; and the stable (critical) fixed point, $(b_+,\delta_+)$, which has only attractive eigenvalues, and should be identified with the inverted XY critical point in our approximation. The two extra fixed points are for $\delta \neq 0$. Both have one attractive and one repulsive eigenvalue. The one for which $\delta$ is positive we denote by $(b_-,\delta_+)$ and the one for which $\delta$ is negative by $(b_+,\delta_-)$. By setting the beta functions equal to zero and solving the resulting equations one is able to find the fixed points for various values of $\lambda$ as shown in Table 1. Indeed, by linearizing our RG equations around these fixed points one is able to calculate the eigenvalues and eigenvectors and show that these coincide with the numerical solutions shown in Fig. 5. Looking at the Table 1 and the Fig. 5, we observe two important trends. For $\bar{\lambda} \neq 0$, the fixed point $(b_+,\delta_+)$ is no longer at $\delta = 0$, but at a positive value of $\delta$. As $\bar{\lambda}$ increases so does this value of $\delta$. We also see that $(b_-,\delta_+)$ starts to move to larger values of $b$. At the value $\bar{\lambda}_c = 2.04$, these two points cohere; and above this they move into the complex plane. For $\lambda = 1$, this corresponds to $N_c \approx 3.4$. We see in Fig. 5c that the effect of losing these two fixed points is to completely destabilize the second order phase transition. We are also able to show numerically that for the runaway trajectories, the coupling that appears in the effective potential with $\langle \boldsymbol{\Phi}_1 \rangle = \langle \boldsymbol{\Phi}_2 \rangle$, $b + \delta/2 \to -\infty$ for $p \to 0$, which should lead to a fluctuation induced first order phase transition.

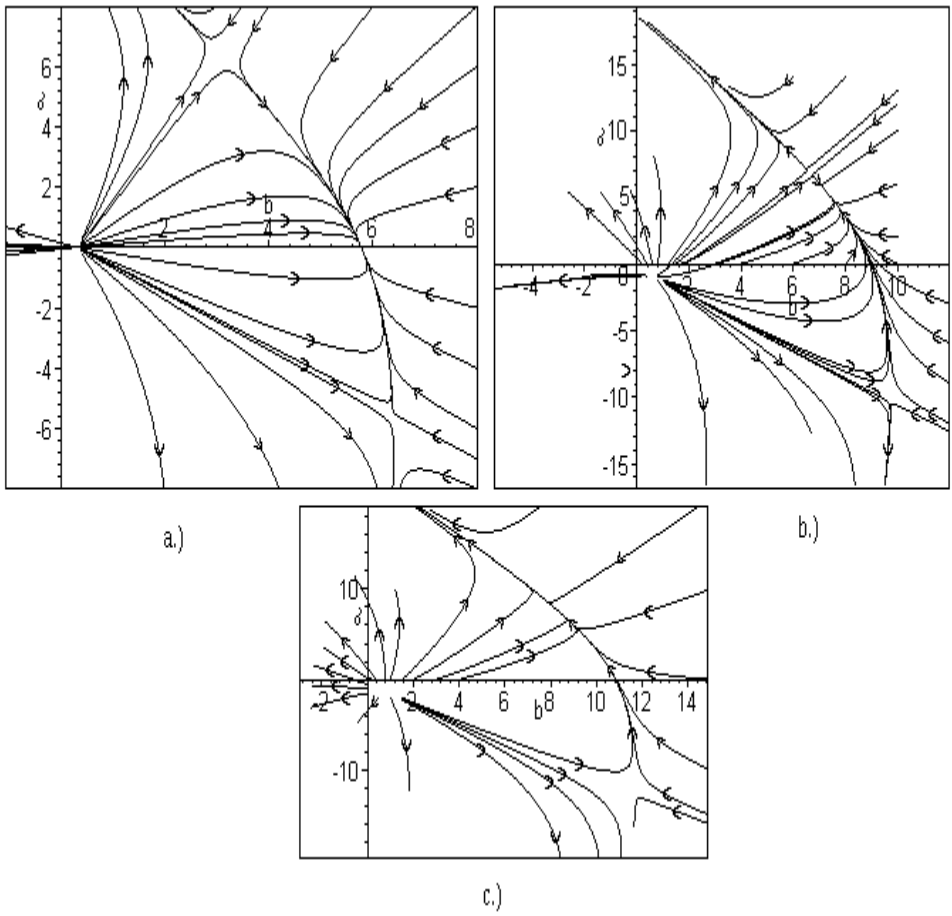

FIG. 5: Flow diagrams in the $(b, \delta)$ plane with initial condition $e = e^*$ for various values of $\bar{\lambda}$. The arrows show the direction of flow as $p \to 0$. a.) $\bar{\lambda} = 0$. b.) $\bar{\lambda} = 2$. c.) $\bar{\lambda} = 2.5$.

In Fig. 6, we plot the separatrices in the space of $(b, e^2)$ between the first and second order phase transitions for selected values of $\bar{\lambda}$. Notice that as $\bar{\lambda}$ increases the boundary between the two phase transitions moves to the right, reducing the domain of attraction of the stable fixed point.

If one would include the $\bar{\lambda}^4$ terms directly into the RG equations, the effect is to more rapidly destabilize the second order phase transition in favour of a first order transition, by pushing down $\bar{\lambda}_c$ to $\bar{\lambda}_c = 1.14$ (corresponding to $N_c = 1.05$ for $\lambda = 1$). Again the same flow structure is present, and $(b_+, \delta_+)$ is destroyed in the same manner as discussed above at $\bar{\lambda}_c$.

We next compute the anomalous dimension $\eta_\Phi$ and the correlation length exponent $\nu$ at the stable fixed point, for $\bar{\lambda} < \bar{\lambda}_c$. These may be obtained by considering $Z_\phi$ and $Z_{\phi^2}$; where the former can be calculated from the composite Greens function $\Gamma^{(2,1)}_{1,1}$, for two separate $\boldsymbol{\Phi}$ fields and a $\boldsymbol{\Phi}^2$ insertion. The terms that contribute to one loop order are shown in Fig. 7. We find

$$\eta_\Phi = -\left(\frac{e^{*2}}{4} + \frac{\bar{\lambda}^2}{\pi^2}\right)$$
$$\nu^{-1} = 2 - \frac{b_+}{8} - \frac{\delta_+}{32} + \frac{e^{*2}}{4} + \frac{\bar{\lambda}^2}{\pi^2} \qquad (12)$$

For $N = 2$ and $\lambda = 1$ this implies $\nu = 0.65$ and $\eta_\Phi = -0.48$. Numerical values of the exponents at other values of $\bar{\lambda}$ are provided in Table. 2. For increasing $\bar{\lambda}$, $\nu$ slightly reduces in value and $\eta$ becomes more negative. Notice that the value of $\nu$ calculated for $\bar{\lambda} = 0$ is very close to the known XY value of $\nu = 0.667$ [28].

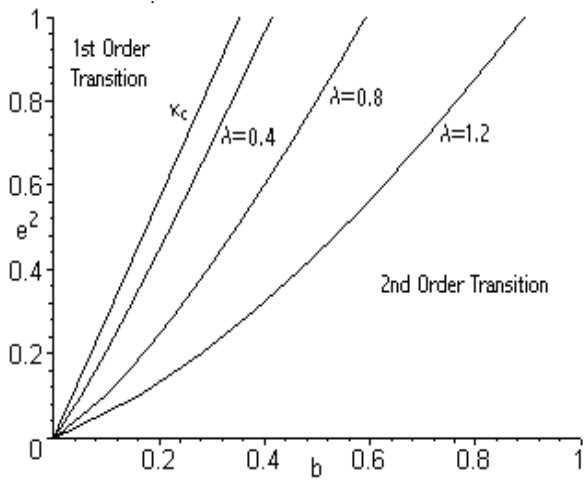

FIG. 6: The boundary in the $e^2 - b$ plane between the fluctuation induced first order phase transition and the second order transition; calculated for $\bar{\lambda} = 0.4$, $\bar{\lambda} = 0.8$ and $\bar{\lambda} = 1.2$. The straight line for $\bar{\lambda} = 0$ is the $\kappa = \kappa_c$ line in the standard HSE [13].

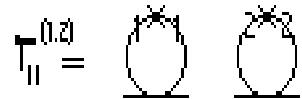

FIG. 7: Terms in $\Gamma^{(1,2)}_{1,1}$ that contribute to $Z_{\Phi^2}$ to one loop order.

| $\bar{\lambda}$ | $\nu$ | $\eta_\Phi$ |
|---|---|---|
| 0 | 0.663 | -0.230 |
| 0.4 | 0.663 | -0.246 |
| 1.0 | 0.659 | -0.331 |
| 1.5 | 0.652 | -0.458 |
| 2 | 0.637 | -0.635 |

TABLE II: Critical exponents $\nu$ and $\eta_\Phi$ calculated for various values of $\lambda$.

## V. RENORMALIZATION GROUP NEAR 4D

When we generalize our theory to arbitary dimension we need to make one modification to the Lagrangian in the Eq. 6, which is to replace the coupling $\bar{\lambda}^2$ by $\epsilon \bar{\lambda}^2$. This is to insure that the theory has a well defined expansion for small $\epsilon = d - 4$, and handles the effect of fermions in a controlled way close to 4D. On integration of the fermion fields we use the well known formula for $\Pi_{\mu\nu}$ for N massless fermion fields, to one loop order, close to 4D

$$\Pi_{\mu\nu} = (p_\mu p_\nu - \delta_{\mu\nu} p^2)\frac{Np^{-\epsilon}}{12\pi^2 \epsilon} + O(1), \qquad (13)$$

to write down a generalised effective theory

$$S_{eff} = \int d^3 x \left[\frac{(\nabla \times \mathbf{A}_-(x))^2}{2\epsilon \bar{\lambda}^2 |\nabla|^{(4-d)}} + \mathcal{L}_h\right], \qquad (14)$$

where close to 4D, $\bar{\lambda}^2 = \lambda^2 N/12$. From the expressions for the loop integrals in $\chi_{\mu\nu}$, $\Gamma^{(2)}$, $\Gamma^{(4)}_{1,1}$, $\Gamma^{(4)}_{1,2}$ and $\Gamma^{(4)}_{2,2}$ given in Appendix A, for arbitrary dimension, it is easy to extract terms singular in $\epsilon = 4 - d$. It should be pointed out that as well as including $\epsilon$-poles we also have to include $\epsilon'$-poles ($\epsilon'$ is defined in the previous section) coming from terms that contain logarithmic singularities in all dimensions, due to the form of the effective $\mathbf{A}_-$ propagator. Since we are using the minimal subtraction scheme, we define $Z_\phi$, $Z_{\phi^2}$, $Z_{A+}$, $Z_{A-}$, $Z_{b1}$ and $Z_{b2}$ so as to subtract out the only $\epsilon$ and $\epsilon'$ poles in the loop integrals. We get the following forms for the renormalized couplings,

$$b_{1,R} p^\epsilon = b_1 - \left[\frac{b_1^2(n+4)}{2} + \frac{b_2^2 n}{2}\right]\frac{p^{-\epsilon}}{8\pi^2 \epsilon}$$
$$+ \left[\frac{3e^2 p^{-\epsilon}}{4\pi^2 \epsilon} + \frac{3\bar{\lambda}^2 p^{-\epsilon'}}{4\pi^2 \epsilon'}\right]b_1 - \frac{6e^4 p^{-\epsilon}}{4\pi^2 \epsilon}$$
$$+ \frac{6\epsilon \bar{\lambda}^4 p^\epsilon}{4\pi^2} + \frac{6e^2 \epsilon \bar{\lambda}^2}{4\pi^2}\frac{p^{-\epsilon'}}{\epsilon'},$$

$$b_{2,R} p^\epsilon = b_2 - \left[ b_2^2 + b_1 b_2 (n+1) \right] \frac{p^{-\epsilon}}{8\pi^2 \epsilon}$$
$$+ \left[ \frac{3e^2 p^{-\epsilon}}{4\pi^2 \epsilon} + \frac{3\epsilon \bar{\lambda}^2 p^{-\epsilon'}}{4\pi^2 \epsilon'} \right] b_2 - \frac{6e^4 p^{-\epsilon}}{4\pi^2 \epsilon}$$
$$+ \frac{6\epsilon \bar{\lambda}^4 p^\epsilon}{4\pi^2} + \frac{6e^2 \epsilon \bar{\lambda}^2}{4\pi^2} \frac{p^{-\epsilon'}}{\epsilon'},$$
$$e_R^2 p^\epsilon = e^2 - \frac{ne^4 p^{-\epsilon}}{12\pi^2}. \quad (15)$$

From the above expressions we derive the following beta functions

$$p \frac{db_1}{dp} = -\epsilon b_1 + \left[ b_1^2 \left( \frac{n'}{4} + 2 \right) + \frac{b_2^2 n'}{4} \right] \frac{1}{8\pi^2}$$
$$- \frac{3(e^2 + \epsilon \bar{\lambda}^2) b_1}{4\pi^2} + \frac{6(e^4 + e^2 \epsilon \bar{\lambda}^2 + \epsilon^2 \bar{\lambda}^4)}{4\pi^2},$$
$$p \frac{db_2}{dp} = -\epsilon b_2 + \left[ b_2^2 + b_1 b_2 \left( \frac{n'}{2} + 1 \right) \right]$$
$$- \frac{3(e^2 + \epsilon \bar{\lambda}^2) b_2}{4\pi^2} + \frac{6(e^4 - e^2 \bar{\lambda}^2 + \epsilon^2 \bar{\lambda}^4)}{4\pi^2},$$
$$p \frac{de^2}{dp} = -\epsilon e^2 + \frac{n' e^4}{24\pi^2}, \quad (16)$$

where $n' = 2n$ is the total number of complex fields in $\mathbf{\Phi}_1$ and $\mathbf{\Phi}_2$. For the two fixed points $(\epsilon b^*, \epsilon \delta^*, \epsilon e^{*2})$, that become the charged fixed points in the standard HSE for $\delta = 0, \lambda = 0$, we find the following equations for small $\bar{\lambda}$

$$b^* = \frac{16\pi^2}{(2n'+8)} \left[ \left( 1 + \frac{3(e^{*2} + \bar{\lambda}^2)}{4\pi^2} - \frac{\delta^* n'}{16\pi^2} \right) \pm \right.$$
$$\left[ \left( 1 + \frac{3(e^{*2} + \bar{\lambda}^2)}{4\pi^2} - \frac{\delta^* n'}{16\pi^2} \right)^2 \right.$$
$$\left. \left. - \frac{3(2n'+8)}{16\pi^2} (e^{*4} + e^{*2} \bar{\lambda}^2) \right]^{1/2} \right],$$
$$\frac{3e^{*2} \bar{\lambda}^2}{\pi^2} = \delta^* \left( \frac{3b^*}{8\pi^2} - 1 \right), \quad (17)$$

where $e^{*2} = \frac{24\pi^2}{n'}$. The fixed points are real for $n' > n'_c$, when it is possible to have a second order phase transition in the theory. $n'_c$ is therefore determined by

$$\left( 1 + \frac{3(e^{*2} + \bar{\lambda}^2)}{4\pi^2} - \frac{\delta^* n'_c}{16\pi^2} \right)^2 - \frac{3(2n'_c + 8)}{16\pi^2} (e^{*4} + e^{*2} \bar{\lambda}^2) = 0, \quad (18)$$

as well as with the Eq. (17). We solve these equations to second order in small $\bar{\lambda}$:

$$n'_c = 182.9 + 126.9 \bar{\lambda}^2 + O(\bar{\lambda}^4). \quad (19)$$

For $\bar{\lambda} = 0$ we recover the well known result of [14]. When small $\bar{\lambda}$ is present $n'_c$ increases. This we take as an indication that again the second order phase transition is destabilized by fermions, in qualitative agreement with the conclusion of the previous section.

## VI. NEUTRAL THEORY IN 3D

It is interesting to see what the coupling to fermions does in the "neutral" case $e = 0$, which would reduce to the XY (not to the inverted XY) model when $\bar{\lambda} = 0$. Such a theory would correspond to the *charged* XY model for the fluctuations of the phase of the superconducting order parameter, which is dual to the Higgs scalar electrodynamics with a *massive* gauge field $\mathbf{A}_+$ [23]. Assuming that such a gauge-field mass is large one can neglect the $\mathbf{A}_+$ field altogether in the Eq. (6), and define what we will call the "neutral" version of the theory.

Setting $\mathbf{A}_+ = 0$ in the Eq. (6) allows one to greatly simplify the theory by making the transformation $\mathbf{\Phi}_2 \to \mathbf{\Phi}_2^*$, upon which the theory becomes symmetric under $\mathbf{\Phi}_1 \longleftrightarrow \mathbf{\Phi}_2$. One may then set $\mathbf{\Phi}_1 = \mathbf{\Phi}_2 = \mathbf{\Phi}$ and $b_1 = b_2 = b$ from the outset, and take $\mathbf{\Phi}$ to have one complex component. This leads to $\mathcal{L}' = \mathcal{L}'_h + \mathcal{L}_f + \mathcal{L}_c$, with

$$\mathcal{L}'_h = |(\nabla - i\mathbf{A}_-) \mathbf{\Phi}|^2 + t_0 |\mathbf{\Phi}|^2 + \frac{b}{4} |\mathbf{\Phi}|^4. \quad (20)$$

A closely related model has been discussed before in [29]. The beta function for the single coupling constant $b$ is readily obtained by taking the $e = 0$ limit of the Eq. (11):

$$p \frac{db}{dp} = -b - \frac{2}{\pi^2} \bar{\lambda}^2 b + (2\sqrt{2} + 1) \frac{b^2}{16}. \quad (21)$$

In contrast to Eqs. (11), now there is always a stable fixed point, at which

$$\eta_\Phi = -\frac{\bar{\lambda}^2}{\pi^2}, \quad (22)$$

$$\nu^{-1} = 2 - \frac{b^*}{8} + \frac{\bar{\lambda}^2}{\pi^2}. \quad (23)$$

For $\lambda = 1$ and $N = 2$ ($\bar{\lambda} = \pi^2/4$) this yields $\eta_\Phi = -0.25$ and $\nu = 0.68$.

There is still, however, a possibility of a discontinuous transition in the theory, since upon integration over $\mathbf{A}_-$ the bare value of coupling $b$ becomes reduced by a contribution $\sim \bar{\lambda}^4$, similarly as in the Eq. (10). If $b$ is small this will turn it negative and the transition should become discontinuous. For large enough $b$, on the other hand, the flow is towards the modified XY critical point with the above exponents.

A related theory of coupled Dirac fermions and scalar field that also exhibits $N$-dependent exponents has recently been considered in [30].

## VII. CONCLUSION AND DISCUSSION

To summarize, we have argued that the coupling of vortex loops to gapless fermionic excitations tends to destabilize the continuous inverted XY transition in favour of a fluctuation induced first-order one. With increase of the coupling to fermions, the domain of attraction of the inverted XY critical point in 3D scalar electrodynamics shrinks, until it finally disappears at some critical value. In our approximation the critical number of fermion components where this happens is $N_c \approx 3.4$, which suggests that for the physical case $N = 2$ the quantum phase transition between dSC and the SDW may still be continuous. We showed that the correlation length exponent decreases with the number of fermions and estimated it to be $\nu \approx 0.65$ for the physical case $N = 2$.

Our number for $N_c$, which happens to be larger than two in our approximation, may well be an overestimate. There are two further effects which can reduce it. First, as we already discussed, retaining $\lambda^4$ terms in the beta function would yield a lower $N_c$. More importantly, tuning our free parameter "c" in the beta functions for HSE to a larger value of the tricritical Ginzburg-Landau parameter would also decrease $N_c$. For example, for $\kappa_c = 0.8/\sqrt{2}$ [31], [32], one finds that $N_c$ is reduced all the way to zero, and the transition becomes discontinuous for all values of the coupling constants.

While there is a great degree of uncertainty in the precise value of the critical number of components $N_c$ inherent to our calculation, the qualitative result that coupling to gapless fermions favours a discontinuous transition, we believe, is likely to be correct. This is because this coupling introduces another massless gauge field into the HSE, the effect of which is, crudely, to make the dual theory effectively more type-I. The reader should recall that, in the phase diagram of standard Higgs scalar electrodynamics [13], increasing the coupling to the gauge-field and making the system more type-I would always eventually lead to the first-order transition. The effect of the additional gauge field is to increase the type-I region in the space of coupling constants at the expense of the type-II, where the transition is continuous. The lowest order calculation suggests that above certain critical value of this coupling, that may be measured by the number of fermionic components $N$, the type-II region disappears completely.

To put our calculation in perspective, it may be useful to ask what the result would be if the superconductor were not gapless, but had a full gap. One can model that situation by adding a mass term $\sim m\bar{\Psi}_j \Psi_j$ to $\mathcal{L}_f$. Integration over fermions would then give an ordinary Maxwell terms for $\mathbf{a}$, which when integrated out would finally yield a *massive* term for $\mathbf{A}_-$, $m\mathbf{A}_{-,t}^2$, where by subscript $t$ we denote the transverse components. Such a massive gauge-field is *irrelevant*, and the transition would remain in the inverted XY universality class. (Similar result is found in the gapless case but at $T \neq 0$, where finite temperature provides an effective mass for the gauge field [11], and makes it irrelevant at the standard KT critical point.) We find it pleasing that in this case one recovers the result that intuitively appears to be correct.

In the previous work [7] the procedure we followed was complementary to the one in this paper: there we integrated approximately the vortex degrees of freedom, to find that when vortices are condensed electrons organise themselves into a SDW. This, however, left the question of the nature of the dSC-SDW transition open. Here by integrating fermions first we study the nature of the phase transition from the vortex point of view. We emphasise, however, that one is still considering one and the same phase transition in the theory (1). We expect the $T = 0$ free energy defined by Eq. (1) to have only one singularity as the parameter $K$ is varied, at which vortex loops blow up (and superconductivity is lost), and simultaneously chiral symmetry for fermions becomes broken (and the SDW is established). This is precisely what happens in the mean-field treatment of the vortex fields, and we believe should be generally true. Our expectation is apparently met in the lattice calculations on a related model of fermions coupled to a bosonic fields via gauge fields [33], where indeed only a single phase transition is found [34].

Theory similar to our Eq. (6) has also been recently studied by Ye [35], but with one important difference: there is only a single scalar field $\mathbf{\Phi}_1$, while $\mathbf{\Phi}_2 = 0$. In such a theory $\mathbf{A}_+$ and $\mathbf{A}_-$ are coupled to each other, and in particular, when $\langle \mathbf{\Phi}_1 \rangle \neq 0$ the integration over $\mathbf{A}_+$ ultimately makes $\mathbf{a}$ *massive* in the insulating phase. As discussed elsewhere [7], this is an artifact of the particular singular gauge transformation made in arriving at the theory. Here we only note that by demanding that $\langle \mathbf{\Phi}_1 \rangle = \langle \mathbf{\Phi}_2 \rangle$, as dictated by the underlying lattice model, completely decouples the gauge fields $\mathbf{A}_+$ and $\mathbf{A}_-$, which then leads to the massless gauge field in the insulating phase as in Eq. (7).

It may be interesting to note a parallel between our model and another popular field theory, namely the Thirring model [36]. Assume that one in the superconducting state, $t_0 > 0$, and integrate all the fields in (6) other than fermions, in the Gaussian approximation. Since in the superconducting state the gauge field $\mathbf{a}$ is massive, one finds a quartic term

$$\sim \frac{\lambda^2}{\sqrt{t_0}} (\bar{\Psi}_j \gamma_\mu \Psi_j)^2$$

in the remaining action for fermions. The theory with just this quartic term would be the 3D Thirring model with $N = 2$. This model has a transition into a phase with broken chiral symmetry as $t_0$ is decreased. While our field theory is not exactly equivalent to Thirring model [37], it is interesting that we find the same trends as observed in its numerical studies [38]: a) the correlation length exponent $\nu$ decreases with $N$, b) above certain $N_c$ the phase transition becomes discontinuous. It would be interesting to see if these non-trivial features of the Thirring model can be obtained in some RG scheme similar to ours.

As applied to underdoped cuprates, our calculation would suggest that the $T \neq 0$ line of the Kosterlitz-Thouless transitions ends in a modified XY critical point with new critical exponents, or even in the first-order transition, at $T = 0$. The latter would imply, for example, that $T = 0$ superfluid density $\rho_{sf}(T = 0)$ may discontinuously drop to zero as doping is reduced, while the former would mean that $\rho_{sf}(T = 0) \sim (x - x_c)^{z\nu}$, where $x_c$ the critical doping, and $z = 1$ the dynamical critical exponent. Detection of these effects is at present probably prohibited due to disorder that is most difficult to control precisely in the underdoped region. In the mean time, numerical calculations on the lattice model (1) should be able to shed some additional light on the problem.

## VIII. ACKNOWLEDGEMENT

This work was supported by NSERC of Canada and Research corporation. We would also like to acknowledge Matthew Case for his kind help with the figures.

## APPENDIX A: EXPRESSIONS FOR VARIOUS 1PI VERTICES TO ONE LOOP ORDER

What follows is a general derivation of various 1PI vertices to one loop order in arbitrary number of dimensions. By applying Feynman rules to the effective theory (described by Eq. (14) in the text) we find the following forms for the polarization tensor

$$\chi_{\mu,\nu}(p) = 2 \int \frac{d^d k}{(2\pi)^d} G(k) - \int \frac{d^d k}{(2\pi)^3}(k_\mu + (k+p)_\mu)G(k)G(k-p)(k_\nu + (k+p)_\nu) \tag{A1}$$

Here, $G(p)$ denotes the bare greens function for both $\mathbf{\Phi}_1$ and $\mathbf{\Phi}_2$ fields. For $\Gamma^{(2)}$ we find the following

$$\Gamma_1^{(2)}(p) = G(p)^{-1} + \left[\frac{b_1(n+1)}{2} + \frac{b_2 n}{2}\right] \int \frac{d^d k}{(2\pi)^d} G(k)$$
$$+ 2e^2 \int \frac{d^d k}{(2\pi)^d} D^+_{\mu,\mu}(k) + 2\bar{\lambda}^2 \int \frac{d^d k}{(2\pi)^d} D^-_{\mu,\mu}(k)$$
$$- e^2 \int \frac{d^d k}{(2\pi)^d} (k+p)_\mu D^+_{\mu,\nu}(k-p)(k+p)_\nu G(k)$$
$$- \epsilon \bar{\lambda}^2 \int \frac{d^d k}{(2\pi)^d} (k+p)_\mu D^+_{\mu,\nu}(k-p)(k+p)_\nu G(k), \tag{A2}$$

where $D^+_{\mu,\mu}(k)$ and $D^-_{\mu,\mu}(k)$ are the bare $\mathbf{A}_-$ and $\mathbf{A}_+$ gauge field propagators, respectively. For $\Gamma^{(4)}_{1,1}$, $\Gamma^{(4)}_{1,2}$ and $\Gamma^{(4)}_{2,2}$ we find

$$\Gamma^{(4)}_{1,1} = \Gamma^{(4)}_{2,2} = b_1 - \left[\frac{b_1^2(n+3) + nb_2^2}{2^{3-d/2}} + \frac{b_1^2}{2}\right]$$
$$\int \frac{d^d k}{(2\pi)^d} G(k) G(k+p)$$

$$-2^{d/2} e^4 \int \frac{d^d k}{(2\pi)^d} D^+_{\mu,\nu}(k+p) D^+_{\mu,\nu}(k)$$
$$-2^{d/2} e^2 \epsilon \bar{\lambda}^2 \int \frac{d^d k}{(2\pi)^d} D^+_{\mu,\nu}(k+p) D^-_{\mu,\nu}(k)$$
$$-2^{d/2} \epsilon^2 \bar{\lambda}^4 \int \frac{d^d k}{(2\pi)^d} D^-_{\mu,\nu}(k+p) D^-_{\mu,\nu}(k),$$
$$\Gamma^{(4)}_{1,2} = \frac{b_2}{2} - \left[\frac{b_2^2}{4} + \frac{b_2^2}{2^{3-d/2}} + \frac{b_1 b_2(n+1)}{2^{2-d/2}}\right]$$
$$\times \int \frac{d^d k}{(2\pi)^d} G(k)G(k+p)$$
$$-2^{d/2} e^4 \int \frac{d^d k}{(2\pi)^d} D^+_{\mu,\nu}(k+p) D^+_{\mu,\nu}(k)$$
$$+2^{d/2} e^2 \epsilon^2 \bar{\lambda}^2 \int \frac{d^d k}{(2\pi)^d} D^+_{\mu,\nu}(k+p) D^-_{\mu,\nu}(k)$$
$$-2^{d/2} \epsilon \bar{\lambda}^4 \int \frac{d^d k}{(2\pi)^d} D^-_{\mu,\nu}(k+p) D^-_{\mu,\nu}(k). \tag{A3}$$

We start by evaluating the following by dimensional regularization

$$2 \int \frac{d^d k}{(2\pi)^d} G(k) -$$
$$\int \frac{d^d k}{(2\pi)^d}(k_\mu + (k+p)_\mu)G(k)G(k-p)(k_\nu + (k+p)_\nu)$$
$$\equiv 2 \int \frac{d^d k}{(2\pi)^d}\left(\frac{1}{k^2} - \frac{(2k_\mu + p_\mu)(2k_\nu + p_\nu)}{k^2(k-p)^2}\right)$$
$$= -\frac{2\Gamma(1-d/2)\Gamma(d/2)^2}{\Gamma(d)(4\pi)^{d/2}}p^{d-4}(p^2 \delta_{\mu\nu} - p_\mu p_\nu),$$
$$\int \frac{d^d k}{(2\pi)^d} G(k)G(k+p) \equiv \int \frac{d^d k}{(2\pi)^d}\frac{1}{k^2}\frac{1}{(k+p)^2}$$
$$= \frac{\Gamma(2-d/2)\Gamma(d/2-1)^2}{\Gamma(d-2)(4\pi)^{d/2}}p^{d-4},$$
$$\int \frac{d^d k}{(2\pi)^d}(k+p)_\mu D^+_{\mu,\nu}(k-p)(k+p)_\nu G(k)$$
$$\equiv 4 \int \frac{d^d k}{(2\pi)^d}\frac{(k^2 p^2 - (k \cdot p)^2)}{k^4(k+p)^2}$$
$$= (d-1)\frac{\Gamma(2-d/2)\Gamma(d/2-1)^2}{\Gamma(d-2)(4\pi)^{d/2}}p^{d-4},$$
$$\int \frac{d^d k}{(2\pi)^d} D^+_{\mu,\nu}(k+p) D^+_{\mu,\nu}(k) \equiv$$
$$\int \frac{d^d k}{(2\pi)^d}\left(\delta_{\mu\nu} - \frac{k_\mu k_\nu}{k^2}\right)\left(\delta_{\mu\nu} - \frac{(k+p)_\mu(k+p)_\nu}{(k+p)^2}\right)$$
$$= \frac{d}{4}(d-1)\frac{\Gamma(2-d/2)\Gamma(d/2-1)^2}{\Gamma(d-2)(4\pi)^{d/2}}p^{d-4}. \tag{A4}$$

Terms involving a single $D^-_{\mu,\nu}$ propagator in both $\chi_{\mu\nu}$ and the $\Gamma^{(4)}$ vertices are logarithmically divergent for all numbers of dimensions, so as discussed in in the text we regularize these terms in by introducing a shift $\epsilon'$ in the scaling dimension of the $\mathbf{A}_-$ propagator. Using this, we write these integrals in the following way

$$\int \frac{d^d k}{(2\pi)^d} (k+p)_\mu D^-_{\mu,\nu}(k-p)(k+p)_\nu G(k)$$
$$\equiv 4 \int \frac{d^d k}{(2\pi)^d} \frac{(k^2 p^2 - (k.p)^2) k^{4-d-\epsilon'}}{k^4 (k+p)^2}$$
$$= \frac{2(d-1)}{\Gamma(d/2)(4\pi)^{d/2}} \frac{p^{-\epsilon'}}{\epsilon'},$$
$$\int \frac{d^d k}{(2\pi)^d} D^-_{\mu,\nu}(k) D^+_{\mu,\nu}(k+p) \equiv$$
$$\int \frac{d^d k}{(2\pi)^d} \left(\delta_{\mu\nu} - \frac{k_\mu k_\nu}{k^2}\right) \left(\delta_{\mu\nu} - \frac{(k+p)_\mu (k+p)_\nu}{(k+p)^2}\right)$$
$$\times \frac{k^{4-d-\epsilon'}}{k^2 (k+p)^2} = \frac{2(d-1)}{(4\pi)^{d/2} \Gamma(d/2)} \frac{p^{-\epsilon'}}{\epsilon'}.$$
(A5)

We also evaluate the $\bar\lambda^4$ term in both $b_1$ and $b_2$

$$\int \frac{d^d k}{(2\pi)^d} D^-_{\mu,\nu}(k) D^-_{\mu,\nu}(k+p) \equiv$$
$$\int \frac{d^d k}{(2\pi)^d} \left(\delta_{\mu\nu} - \frac{k_\mu k_\nu}{k^2}\right) \left(\delta_{\mu\nu} - \frac{(k+p)_\mu (k+p)_\nu}{(k+p)^2}\right)$$
$$\times \frac{k^{4-d} |k+p|^{4-d}}{k^2 (k+p)^2} =$$
$$\frac{(d-1)}{(4\pi)^{d/2}} \left(\frac{\Gamma(d/2-2)}{\Gamma(d/2-1)^2} - \frac{\Gamma(d/2-1)}{2\Gamma(d/2)^2}\right) p^{4-d}.$$
(A6)

By using the RG conditions given in the text (Eq. (9)) and the Ward identities

$$(k-p)_\mu \Gamma^{(3)+,R}(k, k-p, p) = e(\Gamma^{(2)}(k) - \Gamma^{(2)}(p)),$$
$$(k-p)_\mu \Gamma^{(3)-,R}(k, k-p, p) = \lambda(\Gamma^{(2)}(k) - \Gamma^{(2)}(p)),$$
(A7)

one arrives at the Eq. (10) for the renormalized couplings in 3D and for $n = 1/2$.

Close to 4D and at the critical point, we have the following singularities

$$[\chi_{\mu,\nu}(p)]_{sing} = \frac{1}{24\pi^2} \frac{p^{-\epsilon}}{\epsilon} (p^2 \delta_{\mu\nu} - p_\mu p_\nu),$$
$$\left[\Gamma^{(2)}(p)\right]_{sing} = p^2 \left(1 + \frac{3e^2}{8\pi^2} \frac{p^{-\epsilon}}{\epsilon} + \frac{3\epsilon \bar\lambda^2}{8\pi^2} \frac{p^{-\epsilon'}}{\epsilon'}\right),$$
$$\left[\Gamma^{(4)}_{1,1}(p)\right]_{sing} = b_1 - \left[\frac{b_1^2(n+4)}{2} + \frac{b_2^2 n}{2}\right] \frac{p^{-\epsilon}}{8\pi^2 \epsilon}$$
$$- \frac{6e^4 p^{-\epsilon}}{4\pi^2 \epsilon} + \frac{6\epsilon \bar\lambda^4 p^{-\epsilon}}{4\pi^2} - \frac{6\epsilon \bar\lambda^2 e^2 p^{-\epsilon'}}{4\pi^2 \epsilon'},$$
$$\left[\Gamma^{(4)}_{1,2}(p)\right]_{sing} = \frac{b_2}{2} - \left[\frac{b_2^2}{2} + \frac{b_1 b_2(n+1)}{2}\right] \frac{p^{-\epsilon}}{8\pi^2 \epsilon}$$
$$- \frac{6e^4 p^{-\epsilon}}{4\pi^2 \epsilon} + \frac{6\epsilon \bar\lambda^4 p^{-\epsilon}}{4\pi^2} + \frac{6\epsilon \bar\lambda^2 e^2 p^{-\epsilon'}}{4\pi^2 \epsilon'}.$$

From these it is straightforward to define $Z_\phi$, $Z_{\phi^2}$, $Z_{A+}$, $Z_{A-}$, $Z_{b1}$ and $Z_{b2}$, that remove these singular terms. Using the relations between renormalized and bare couplings $e_R^2 = p^{-\epsilon} Z_{A+} e^2$, $\bar\lambda_R^2 = Z_{A-} \bar\lambda^2$ $b_{1R} = p^{-\epsilon} Z_\phi^2 Z_{b1}^{-1} b_1$ and $b_{2R} = p^{-\epsilon} Z_\phi^2 Z_{b2}^{-1} b_2$ we arrive at the Eqs. (15) given in the text.

---